\documentclass[12pt,preprint]{aastex}

\newcommand{\hst}{{\sl HST}}
\newcommand{\omcen}{{$\omega$ Cen}}

\slugcomment{Accepted version, 30 Nov.\ 2004 }

\shorttitle{Metallicity of \omcen\ main sequence}
\shortauthors{Piotto et al.}

\begin{document}


\title{Metallicities on the Double Main Sequence of $\omega$ Centauri\\
Imply Large Helium Enhancement\footnote{Based on
observations with the ESO {\it Very Large Telescope + FLAMES}, at the
Paranal Observatory, Chile, under DDT program 272.D-5065.}}

\author{Giampaolo Piotto\altaffilmark{2},
             Sandro Villanova\altaffilmark{2},
             Luigi R. Bedin\altaffilmark{2,3},
             Raffaele Gratton\altaffilmark{4},
     	Santi Cassisi\altaffilmark{5},
     	Yazan Momany\altaffilmark{2},
     	Alejandra Recio-Blanco\altaffilmark{2},
     	Sara Lucatello\altaffilmark{4},
     	Jay Anderson\altaffilmark{6},
     	Ivan  R. King\altaffilmark{7},
     	Adriano Pietrinferni\altaffilmark{5}, and
     	Giovanni Carraro\altaffilmark{2}}

\altaffiltext{2}{Dipartimento  di Astronomia, Universit\`a  di Padova,
Vicolo dell'Osservatorio 2, I-35122 Padova, Italy;
piotto-villanova-momany-recio-carraro@pd.astro.it}

\altaffiltext{3}{European             Southern            Observatory,
Karl-Schwarzschild-Str.\  2, D-85748 Garching, Germany; lbedin@eso.org}

\altaffiltext{4}{INAF-Osservatorio Astronomico di Padova, Vicolo
       dell'Osservatorio 5, I-35122 Padova, Italy;
gratton-lucatello@pd.astro.it}

\altaffiltext{5}{INAF-Osservatorio Astronomico di Collurania,
via M. Maggini, 64100 Teramo, Italy
cassisi-adriano@te.astro.it}

\altaffiltext{6}{Department of  Physics and Astronomy,  Mail Stop 108,
Rice    University,   6100   Main    Street,   Houston,    TX   77005;
jay@eeyore.rice.edu}

\altaffiltext{7}{Department of Astronomy, University of Washington,
Box 351580, Seattle, WA 98195-1580; king@astro.washington.edu}


\begin{abstract}
Having shown in a recent paper that the main sequence of $\omega$
Centauri is split into two distinct branches, we now present
spectroscopic results showing that the bluer sequence is {\it less}
metal-poor.  We have carefully combined GIRAFFE@VLT spectra of 17
stars on each side of the split into a single spectrum for each
branch, with adequate $S/N$ to show clearly that the stars of the blue
main sequence are less metal poor by 0.3 dex than those of the dominant
red one. 
From an analysis of the individual spectra, we could not 
detect any abundance spread among the blue main sequence stars,
while the red main sequence stars show a 0.2 dex spread in metallicity.
We use stellar-structure models to show that only greatly
enhanced helium can explain the color difference between the two main
sequences, and we discuss ways in which this enhancement could have
arisen.
\end{abstract}



\keywords{globular    clusters:    individual (\objectname[\omecen]{NGC
         5139}) --- Hertzsprung-Russell diagram --- Galaxy: abundances}


%
\section{Introduction}
%
Omega Centauri  is the  Galactic globular cluster  (GC) with  the most
complex stellar  population, and a  huge amount of attention  has been
paid to it.  Its high mass may represent a link between GCs and larger
stellar systems.  Understanding the star-formation  history in \omcen\
might give fundamental information  on the star-formation processes in
more complex systems such as galaxies.

The problem  is that the more  information we acquire on  the stars of
\omcen, the less we understand their origin. In this respect, the most
recent  and most  puzzling result  surely is  the identification  of a
double main sequence (DMS)---originally discovered by Anderson (1997),
and discussed in detail by Bedin et al.\ (2004a, B04), who showed that
the feature is real, and present throughout the cluster.  As far as we
know, \omcen\  is the only  GC to show  two MS populations.   But what
makes this result even more enigmatic is the color distribution of the
stars on the  DMS.  If we were  to guess what the MS  should look like
from what the spectroscopic (Norris and Da Costa 1995) and photometric
(Hilker and  Richtler 2000,  Lee et al.\  1999, Pancino et  al.\ 2000)
investigations  of the stars  on the  giant branch  tell us,  we would
expect it to  show a concentration to a blue  edge, corresponding to a
metal-poor ([Fe/H]  $\sim-1.6$) population containing the  bulk of the
stars,   a   second,   less    blue,   group   corresponding   to   an
intermediate-metallicity  population  ([Fe/H] $\sim-1.2$),  containing
about 15$\%$  of the stars  (according to Norris, Freeman,  \& Mighell
1996), and  a small, even  redder component from a  metal-rich ([Fe/H]
$\sim-0.5$) population with 5$\%$ of the stars (Pancino et al.\ 2000).
The sequence shown in Fig.~1 of  B04 (see \ also Fig.~7 of this paper)
could  not be  more different  from these  expectations:  The observed
sequences are clearly  separated, and the bluer MS  (bMS) is much less
populous  than the  red MS  (rMS), containing  only $\sim$25\%  of the
stars.

B04 discussed a  number of possible explanations:\ that  the bMS could
represent (i)  a super-metal-poor  ([Fe/H] $\ll -2.0$)  population, or
(ii) a super-helium-rich  ($Y>0.3$) population (a hypothesis exploited
further by Norris 2004), or,  finally, (iii) a background object, 1--2
kpc beyond  \omcen.  This paper provides essential  information on the
metal content  of the two  sequences, showing a possible  path through
the morass of contradictory results outlined above.

%
\section{Observations and Data Reduction}
%
We  observed  \omcen\  on  ESO   DDT  time  in  April--May  2004  with
FLAMES\-@VLT+GI\-RAFFE,  under  photometric   conditions  and  with  a
typical seeing of  0.8 arcsec.  We used the  MEDUSA mode, which allows
obtaining  130 spectra simultaneously.   To have  enough $S/N$  on the
faint main-sequence  stars, and to cover the  wavelengths of interest,
we  used the  low-resolution mode  LR2,  which gives  $R=6400$ in  the
3960--4560~\AA\ range. Twelve one-hour  spectra were obtained for each
of 17 rMS  stars and 17 bMS stars (in  the magnitude range $20<V<21$),
selected from the \hst\ ACS field 17 arcmin southwest of the center of
\omcen, shown in Fig.~1d of B04.  The selected MS stars are plotted in
the  color-magnitude diagram  of  Fig.~7.  The  remaining fibers  were
pointed at 88  subgiant branch (SGB) stars equally  distributed on the
three  SGBs of  Fig.~1b in  B04.   This paper  presents a  preliminary
analysis of the MS spectra.  Table~1 lists their coordinates and their
magnitudes and colors in the ACS F606W, F606W $-$ F814W system.

The data were reduced using the GIRAFFE pipeline, in which the spectra
have  been corrected  for bias  and  flat field,  and then  wavelength
calibrated   using  both   prior  and   simultaneous  calibration-lamp
spectra.  The resulting  spectra have  a dispersion  of 0.2~\AA/pixel.
Then  each   spectrum  was   corrected  for  its   fiber  transmission
coefficient,  obtained by measuring  for each  fiber the  average flux
relative to a  reference fiber, in five flat-field  images. Finally, a
sky correction was applied to each stellar spectrum by subtracting the
average of ten sky spectra observed simultaneously (same FLAMES plate)
with the star.   The final MS single spectra have  a typical $S/N \sim
2$--3.

In order  to increase  the $S/N$, for  each MS  star we summed  all 12
one-hour spectra.   Although the spectra  were taken over a  period of
time, we ignored differences  in heliocentric correction, because they
are far too small to matter.  The brightest among the resulting 17 bMS
stacked spectra was  cross-correlated with each of the  others, to get
the differential radial velocities. The same operations were performed
with the 17 rMS stacked spectra. We found $\langle v_{\rm rad} \rangle
= 232 \pm  2$ km s$^{-1}$ for the bMS stars,  and $\langle v_{\rm rad}
\rangle = 235 \pm 3$ km s$^{-1}$ for the rMS. It is noteworthy, by the
way, that  the stars  in both sequences  have the same  average radial
velocity.  (See discussion in \S 5.)

Finally, the spectra were shifted  and co-added in order to obtain one
single bMS  and a  single rMS spectrum,  for a  total of 204  hours of
exposure time per spectrum, and an average $S/N\sim30$.

%

\section{Abundance measurement}

\subsection{Average metallicities of the bMS and rMS}

The effective temperatures that we  used are the average of the values
derived from the star colors and from the profiles of H$_\gamma$.  For
the color temperatures we used F606W $-$ F814W colors derived from the
\hst\ ACS  color-magnitude diagram  (CMD) of B04,  along with  our own
evolutionary models (see below), in both cases using the prescriptions
given  by Bedin  et  al.\  (2004b) for  transformations  into the  ACS
observational plane.  The individual bMS  stars cover a range of about
370  K, and  the  rMS stars  span  a temperature  interval  of 330  K.
Temperatures derived from individual  stars were averaged by weighting
stars  according to their  flux in  the $B$-band,  which approximately
covers  the wavelength region  of the  spectra. The  temperatures from
H$_\gamma$\  were found  by  comparison with  synthetic profiles  (see
Fig.~1) obtained  using the Kurucz  (1992) model atmospheres  (with no
overshooting) and  the prescriptions by Castelli et  al.\ (1997).  The
temperatures derived  from H$_\gamma$  are somewhat higher  than those
from colors (by about 200~K);  however, we regard this disagreement as
being  within  the uncertainties  in  both  determinations. The  final
adopted  temperatures are  5200~K  for  the rMS,  and  5400~K for  the
bMS.  Uncertainties in  these temperatures  are $\pm  100$~K,  but the
difference is better determined.


For the surface gravities, we adopted  the usual value for MS stars of
$\log g=4.5$ for both sequences. While the bMS is fainter by about 0.3
mag  (implying  smaller radii  by  about 6\%),  there  may  be a  mass
difference of  about 15\%  between the two  sequences, if  they indeed
differ in their He content (as argued in \S 5), roughly offsetting the
difference in radii.  Lacking  information, we used a micro-turbulence
velocity of 1 km s$^{-1}$ for both sequences, roughly the solar value.

The  model atmospheres  of Kurucz  (1992) used  throughout  this paper
assume $N_{\rm He}/N_{\rm H}=0.1$ by number, corresponding to $Y=0.28$
by  mass.  Detailed calculations  using  appropriate model  atmosphere
codes  are needed  to  take properly  into  account the  impact of  He
abundances strongly  different from this  value, as suggested  in this
paper for the  bMS stars.  As a preliminary  exploration, we estimated
the  possible magnitude  of this  effect by  using a  simplified model
atmosphere  code, in  which the  run of  the temperature  with optical
depth is  not modified,  and it  is assumed that  opacity is  a simple
function of temperature and  electron pressure.  Two model atmospheres
were  computed  under  these  assumptions,  adopting  the  temperature
stratification of  the original Kurucz model,  with $N_{\rm He}/N_{\rm
H}=0.1$ and 0.3 by number, respectively. These correspond to $Y=0.28$\
and 0.54  by mass,  respectively. The two  model atmospheres  are very
similar,  but the  He-rich atmosphere  has a  slightly  lower electron
pressure,  by about  10--15\%.  This  is because,  due  to the  larger
molecular  weight,  the  ratio   between  electron  and  gas  pressure
decreases slightly when the He content is increased. The net effect of
this is to decrease by the same amount the H$^-$ opacity, which is the
dominant opacity  source in the atmospheres of  these stars.  Reducing
the  continuum opacity  makes the  atmospheres more  transparent, thus
enhancing  line strength.  In practice,  neglecting this  effect would
cause us  to overestimate abundances by 10--15\%  (that is, 0.04--0.06
dex) in He-rich stars. Note, however, that the He abundance difference
required to  explain the  difference between the  bMS and rMS  is only
half of the  change we assumed in our  exploratory computations. Hence
we expect  the effect  to be  about half that  quoted above  (that is,
0.02--0.03 dex).

Metal abundances  (mainly Fe) were obtained by  comparing the observed
average  spectra  for the  bMS  and  the  rMS with  synthetic  spectra
computed with  different metal abundances.  We adopted a  single model
atmosphere in our computations of  synthetic spectra for all bMS stars
and a single one for all  rMS stars. We tested that this approximation
does  not introduce  significant  errors by  performing the  following
exercise. We computed  synthetic spectra with temperatures appropriate
for  each individual  star, and  then averaged  them for  bMS  and rMS
sequences using the same weight criterion (luminosity) used to average
the  temperatures. We  then compared  this average  synthetic spectrum
with  a   synthetic  spectrum  computed  with   the  weighted  average
temperature. The  two spectra  are almost indistinguishable  from each
other:  the  largest  intensity   differences  are  at  the  level  of
0.001. This  implies  differences  of  less  than  0.004  dex  in  the
abundances. This  possible source of error is  clearly negligible with
respect to other sources of error.

The  region 4405--4445~\AA\  was selected  for the  comparison  of the
observed and synthetic spectra,  because it contains numerous metallic
lines (mainly  due to Fe-peak  elements, with a  few strong Ca  and Ti
lines), but  few lines due to molecules  (CH and CN), and  no strong H
lines, as shown  by the upper panel of Fig.~2,  where we over-plot the
average bMS  and rMS spectra.  The synthetic spectra were  smoothed to
the resolution of the observed spectra. Our best values are those that
minimize the r.m.s.\ scatter of the residuals between the observed and
synthetic  spectra.  Using a  solar spectrum  from the  literature, we
verified   that  this  procedure   accurately  reproduces   the  solar
abundance. The results are shown  in Fig.~2 (middle and lower panels);
the best values are [M/H] $=-1.57$ for the rMS, and [M/H] $=-1.26$ for
the  bMS. Internal  errors of  $\pm0.1$~dex in  these  abundances were
estimated  by comparing  the values  obtained  from each  half of  the
wavelength  range  separately.  Systematic  errors  are  dominated  by
uncertainties in the  adopted temperatures:\ a change of  100~K in the
effective  temperatures  causes a  change  of  $\sim  0.2$~dex in  the
derived  abundances.  Note  that the  temperature  uncertainty affects
mainly the absolute  metallicities, and has much less  effect on their
difference.

%
%

Fig.~2 clearly shows  that the bMS stars are more  metal rich than the
rMS ones.   (Note that the differences  are made less  apparent by the
fact that the bMS stars have a higher temperature.)  The [Fe/H] of the
rMS is consistent  with the peak of the  abundance distribution of red
giants in \omcen\ (Suntzeff \& Kraft 1996; Norris et al.\ 1996), while
the bMS  corresponds roughly  to the second  peak in  the distribution
obtained by those  authors. Also, the relative number  of stars in the
two sequences is  consistent with the relative numbers  in the RGB, as
noted by B04.

%
%

\subsection{C, N, and Ba abundances}

Carbon abundances were obtained  by comparing the averaged spectra for
the red and blue main sequences with synthetic spectra (Fig.~3) of the
spectral region 4300--4330~\AA, including the $\Delta v=0$ strong band
heads  of the  A$^2\Delta$-X$^2\Pi$\ transition  of CH,  computed with
appropriate model-atmosphere parameters and  different values of the C
abundances.  From this comparison, we  found values of [C/M] = 0.0 for
both sequences.

%
%

Nitrogen abundances  were found by  a similar comparison  (Fig.~4) for
the region  4200--4225~\AA, including the $\Delta v=-1$  band heads of
the X$^2\Sigma$-B$^2\Sigma$\  CN transition.  A rather large  value of
[N/M] $\sim  1.0$\ or 1.5 was found  for the bMS. The  N abundance for
the  rMS  is not  well  constrained: values  of  [N/M]  $\leq 1$\  are
compatible  with  observations.    Finally,  the  Ba  abundances  were
obtained from the resonance line of Ba~II at 4554~\AA.  Our values are
[Ba/M] = +0.7 and +0.4 for the bMS and rMS, respectively (Fig.~5).

%
%

These comparisons show that the bMS is not very rich in C; hence these
stars cannot have been formed from  the ejecta of C stars. This result
becomes relevant when  we try to interpret the  origin of the chemical
anomalies  implied by  the presence  of the  bMS (see  \S 6).   The Ba
abundance for  stars on  the bMS is  compatible with  (albeit somewhat
smaller than) that observed in  metal-rich red giants of \omcen\ (cf.\
Smith et al.\ 2000).

\section{Individual Spectra and Abundance Spreads}

While  the  $S/N$\ of  the  summed  spectra  for individual  stars  is
generally low ($\leq$ 10 per pixel along the direction of dispersion),
they  may  provide  useful  information  on  the  composition  of  the
individual stars. The procedure  we followed was the following. First,
we re-binned  the spectra at a  resolution of 2~\AA\ per  bin. At this
resolution the typical  $S/N$ of the spectra is  now $\sim 30$, enough
for  line-index measurements.  Second,  we measured  mean instrumental
intensities within  a number of  narrow spectral bands  (see Table~2),
centered on  strong spectral absorption features  as well as  in a few
reference  ``pseudo-continuum"  spectral  ranges.  We then  derived  a
number of spectral indices by dividing the instrumental mean intensity
measured in the bands containing  the features by the weighted average
of adjacent  ``pseudo-continuum" bands. The weights were  given by the
distances  (in wavelength) between  the ``pseudo-continuum"  bands and
the  bands containing the  absorption features.  Error bars  for these
indices were obtained by considering  the $S/N$\ of the spectra at the
wavelength  of   each  band,  the  bandwidths,  and   by  summing  the
contributions   to  errors  of   both  line   and  ``pseudo-continuum"
bands.  Typical internal errors  are 0.059,  0.048, 0.028,  0.030, and
0.029 for  H$_\delta$, Ca~I~4227,  G band, H$_\gamma$,  and Fe~I~4383,
respectively.

%
%

Figure~6  displays  the  runs   of  some  of  these  spectral  indices
(Fe~I~4383, Ca~I~4227, and G-band) with  the F606W $-$ F814W color for
the program stars. Different symbols are used for stars of the bMS and
rMS.

The sequences for the bMS stars  are nearly as narrow as expected from
the internal errors in colors  and spectral indices, with the possible
exception only of  star 3348, which may have a strong  G band: this is
shown well by Table~3, which compares the spread in line indices (at a
given  color) expected from  internal errors  (in colors  and spectral
indices) with  the observed  r.m.s.\ around the  best-fitting straight
line.  The values  listed  in this  table  indicate that  star-to-star
abundance  variations within  bMS  stars have  not  been detected.  By
comparing the observed spread  with the temperature sensitivity of the
spectral features, we may roughly estimate that star-to-star abundance
variations  within bMS stars  are well  below 0.08~dex  (r.m.s.), with
similar values provided by all three indices. The very small abundance
spreads for  bMS stars that are  indicated by line  indices agree well
with  the very  small width  of the  sequence in  the  color magnitude
diagram.  We conclude  that the  bMS is  populated by  stars  having a
fairly  uniform  chemical   composition.  This  might  have  important
implications on its origin.

On the other hand, the  rMS stars display star-to-star scatters in the
line  indices that are  much larger  than expected  from observational
errors  alone, and  larger  than  those obtained  for  bMS stars  (see
Table~3).  Note   that  we  omitted  from  these   plots  and  further
comparisons three  stars (237, 664,  and 3222) having spectra  of very
low  quality. The  large  spreads are  obtained  consistently for  all
metallic spectral indices for the  rMS stars. If we compare the excess
spread with the expectations based  on internal errors, we may roughly
estimate that  there are  star-to-star abundance variations  among rMS
stars of  about 0.15--0.20~dex  (r.m.s.\ scatter), again  with similar
values provided by all three indices. Such a spread is consistent with
the observed width in color of the rMS (about 0.02 mag), which is much
larger than that of the bMS (0.008 mag). This value is also consistent
with the spread in chemical composition usually found for $\omega$~Cen
stars (Suntzeff \&  Kraft 1996; Norris, Freeman \&  Mighell 1996), and
is therefore not unexpected.

\section{Interpreting the observational results}

\subsection{A super-helium-rich population?}

The main result of our investigation  is surely the fact that {\it the
bMS is $0.3\pm  0.2$~dex more metal rich than the  rMS}, the error bar
being essentially  due to uncertainties in  the relative temperatures.
This definitely  removes the  possibility raised in  B04 that  the bMS
represents a super-metal-poor population.

The second piece  of evidence provided by our  spectra, i.e., that the
bMS and  rMS have the same  radial velocity, makes  the already remote
possibility  that the  bMS represents  a background  object  even more
unlikely. In addition, B04 have shown that the DMS is present from the
cluster center to at least 17 arcmin from the center. We also verified
that the WFPC2 field at 7  arcmin (Fig.~1c in B04) from the center and
the outer ACS field (Fig.~1d in B04) have approximately the same ratio
of bMS to rMS stars. Finally, we note that preliminary results by some
of us (Anderson 2003, Anderson \& King 2005, in preparation) show that
the   average   proper   motions    of   the   two   populations   are
indistinguishable.

In summary,  all observational evidence  indicates that the  bMS stars
are \omcen\ members. Still,  we remain with the puzzling observational
result that the  bMS stars are more metal rich than  the rMS ones. The
problem is  that any canonical stellar models  with canonical chemical
abundance tell us that the blue  MS should be more metal poor than the
red MS.

One of the hypotheses made  by B04, and further investigated by Norris
(2004) is  that   the  bMS  might   have  a  strong   He  enhancement.
Interestingly  enough,  from  his  theoretical  investigation,  Norris
(2004) supposed a  metal content for the  bMS and rMS  very similar to
the  values that  we  have  measured.  In  view  of our  observational
findings, the He overabundance seems to be the only way to explain the
MS split.  Though  a direct measurement of He for the  MS stars is not
feasible, we are in a position to test this hypothesis, as we now know
the metallicity of both the bMS and the rMS.

%
%

Because of the sizable dependence of the CMD location of the MS on the
He content, we tried to make an indirect estimate of the He content of
the bMS by  comparison with suitable stellar models.  We used the most
updated physics to calculate  specific models for the metallicities of
the two  sequences.  We refer to  Pietrinferni et al.\  (2004) for the
details  of the  models, as  well  as the  adopted physical  scenario.
Unlike Pietrinferni  et al.\ (2004),  however, in the present  work we
assumed  [$\alpha$/Fe]  =  0.4;  the  alpha-enhanced  models  will  be
published  in a  forthcoming paper.   As  for the  adopted initial  He
content, we assumed  a canonical $Y=0.246$ (Salaris et  al.\ 2004) for
the  $Z=10^{-3}$ metallicity  appropriate for  the rMS,  and  we tried
$Y=0.248$,    $Y=0.35$,    and    $Y=0.45$   for    the    metallicity
$Z=2\times10^{-3}$ appropriate for the bMS. The results of the fit are
shown  in Fig.~7.  Clearly, the  model  with the  standard He  content
cannot fit  the bMS. {\it The  bMS can be reproduced  only by adopting
$0.35<Y<0.45$}.  The  ridge line  of  the bMS  is  best  fitted by  an
isochrone for $Y=0.38$.  Our conclusions  are very similar to those of
Norris (2004), but they are based on the measured MS metallicities and
on more up-to-date stellar models.

\subsection{The progeny of the bMS}

No  such  high He  abundance  has  ever been  found  in  any other  GC
(Cassisi, Salaris, \& Irwin 2003,  Salaris et al.\ 2004), though there
are  a few  anomalous clusters  that might  have He  overabundance, as
discussed  at the  end of  this Section.   Surely, it  is not  easy to
explain  its  origin  (see  next  Section).  Therefore,  even  if  our
hypothesis seems robust, we searched for other, independent evidence.

In particular, we looked for the  progeny of the bMS He-rich stars. We
do not expect  to learn much from the red  giant branch (RGB), because
of  its spread  in metallicity,  but  we do  expect that  a star  with
$Y>0.35$  will  reach the  HB  with a  smaller  mass,  because of  the
combined  effect of the  decrease of  the evolving  mass (evolutionary
lifetimes  decrease strongly when  the He  content increases)  and the
increase of total mass loss due  to the longer RGB lifetime.  Its zero
age HB location  should therefore be on average  hotter (bluer) than a
HB star with canonical He  content (D'Antona et al.\ 2002, D'Antona \&
Caloi 2004;  see also Sweigart  1997). More specifically,  our stellar
models  predict,   for  a  fixed   stellar  age  and   mean  mass-loss
efficiency\footnote{Note also that the mass-loss efficiency might even
increase in  He-overabundant stars, because  of the smaller  masses on
the RGB.}, that the mean ZAHB effective temperature increases by $\sim
1500$ K when the He abundance  is raised from $Y=0.25$ to 0.35, but it
increases  by  $\sim15,000$  K  when  the He  abundance  changes  from
$Y=0.25$ to $\sim0.45$.

Indeed, \omcen\ does have an anomalously blue HB (D'Cruz et al.\ 2000,
Momany  et al.\  2004), extending  to a  temperature  corresponding to
stars  which must  have almost  completely lost  their  envelope.  The
extended  HB  of  \omcen\  shows  a  clear  discontinuity  at  $T_{\rm
eff}\sim20,000$~K,  as is  clearly visible  in the  $U$, $U-V$  CMD of
Fig.~2 in Momany et al.  Also D'Cruz et al.\ find a discontinuity at a
similar temperature (gap G3 in their notation), in a far-UV CMD.  From
the CMD  of Momany et al.,  we estimate that  at least 17\% of  the HB
stars  in   \omcen\  are   hotter  than  the   $T_{\rm  eff}=20,000$~K
discontinuity (hereafter extremely-hot HB  stars, EHB). The counts are
not complete in  the faint (hotter) part of the HB  in that $U$, $U-V$
CMD. On the other hand, D'Cruz et al.\ estimate that 32\% of the stars
are hotter than the same  discontinuity.  In their diagram these stars
are the  brightest stars, while some incompleteness  is surely present
in the counts of the cooler  (fainter) stars in their F160W, F160W $-$
$V$ CMD.  In conclusion, we estimate that between 20\% and 30\% of the
HB  stars of  \omcen\ are  hotter  than the  discontinuity located  at
$T_{\rm eff}\sim20,000$~K.  Interestingly enough,  in the ACS field at
17 arcmin from the center, we estimate that $24\pm3$\% of the MS stars
are in the  bMS.  It is therefore tempting to  associate the EHB stars
in \omcen\ with the progeny  of the bMS, He-rich stars.  Norris (2004)
also came to a similar conclusion.  The EHB stars need not necessarily
include all the  progeny of the bMS, because of  the dispersion in the
mass loss during  the RGB phase, but the star  counts and the presence
of the  discontinuity seem to indicate  that most of them  could be on
the EHB.  It  might also be relevant that Moehler  et al.\ (2002) have
found  the  hottest  HB  stars  ($T_{\rm eff}>30,000$  K)  to  have  a
supra-solar  He content.  Moehler et  al.\ interpret  this anomalously
high He abundance  as the effect of the extra  mixing in late He-flash
stars (Brown  et al.\ 2001).  If  our hypothesis is  correct, at least
part of the He enhancement could  be primordial.  An easy test of this
idea would  be the  measurement of  the He content  for HB  stars with
$T_{\rm eff}<30,000$,  which should  not experience the  extra mixing,
according  to the  Brown  et  al.\ (2001)  models.   We are  presently
investigating  whether other  EHB GCs  show  evidence of  a double  or
broadened MS.

The He  content also affects  the HB luminosity,  in the sense  that a
higher helium  content would  imply a brighter  HB.  In  this respect,
Norris (2004)  pointed out that a  higher helium content  for the more
metal-rich population in  the bMS of \omcen\ seems  to be contradicted
by the  results of Butler,  Dickens, \& Epps  (1978), and Rey  et al.\
(2000), who find  that the RR Lyraes with  $-1.3<$[Fe/H]$<-1.0$ are on
the  average  fainter  than  the  more  metal-poor  ones  by  0.2--0.3
magnitude (though there are at  least two very bright, likely evolved,
metal-rich RR  Lyraes). However, as discussed above,  it is reasonable
to assume that  if the HB stars of the  metal-rich population are very
rich in  He, most of them  must be hotter than  the instability strip,
and therefore, as  pointed out by Norris (2004),  the RR Lyraes cannot
be  representative  of the  entire  cluster  population.   Hints of  a
brighter HB  than expected  (from models) for  \omcen\ are  visible in
Fig.~2 of  Momany et  al.\ (2004),  where the HB  of OmCen  from $\sim
6,000$ K to $>30,000$ K is  shown.  While the HB luminosity of \omcen\
needs further investigation, it is surely interesting to note that two
of the  most massive  GCs of our  Galaxy, NGC~6388 and  NGC~6441, both
have HBs that are still enigmatic,  both because of the presence of an
extended  blue tail (Rich  et al.\  1997) despite  their metallicities
([Fe/H]  = $-0.6$  and  $-0.5$, respectively),  and  because they  are
populated  by  anomalously bright  RR  Lyraes  (Pritzl  et al.\  2002,
2003). Moreover,  these two clusters have  a tilted HB,  with the blue
side brighter  than the red one  (Rich et al.\ 1997,  Raimondo et al.\
2002). Interestingly  enough, Sweigart \& Catelan (1998),  in order to
explain the HB  anomalies of NGC~6388 and NGC~6441,  proposed a helium
value $Y=0.38$--0.43  (very similar to the helium  enhancement we need
to explain  the bMS of  \omcen).  Also, M13  is well known to  have an
anomalously bright and extended HB,  brighter than the HB of a cluster
with very  similar metallicity  like M3. Johnson  \& Bolte  (1998) and
Paltrinieri et  al.\ (1998)  have shown that  the anomalous HB  of M13
cannot  be  due to  an  age effect.  Johnson  \&  Bolte interpret  the
difference between the  HBs of M13 and M3 as due  to a higher ($\Delta
Y\sim0.05$) helium  content in M13.  There are a number  of (indirect)
evidences in the  literature that there could be  populations of stars
with an  enhanced helium content in  GCs, and this effect  needs to be
investigated in more detail.

\section{Discussion}

If an extremely high helium content is the explanation of the abnormal
bMS of \omcen,  the immediate question that arises  is: Where does all
this He  come from? Most authors  try to explain  the abundance spread
within  \omcen\ as  a  peculiar chemical  evolution  history for  this
object, which possibly  was once the nucleus of a  dwarf galaxy (for a
comprehensive discussion of the  literature, see Gratton, Carretta, \&
Sneden  2004). In  this framework,  the enormous  production of  He is
attributed  to pollution  by the  ejecta  of a  well-defined group  of
stars.  If  we compare the bMS  and rMS abundances,  the difference of
helium  abundance  required  to  explain  our data  is  about  $\Delta
Y=0.14$,  while the  analogous  variation in  heavy  metal content  is
$\Delta  Z<0.002$ (assuming that  the variation  of abundances  of the
Fe-peak  elements  is  representative  of  all  metals).  The  $\Delta
Y/\Delta  Z>70$ suggested  by  these data  is  more than  an order  of
magnitude  larger  than  the  value  found  appropriate  for  Galactic
chemical evolution (see, e.g., Jimenez  et al.\ 2003).  It is possible
that  the mass  of  \omcen\ was  just  right to  allow  the ejecta  of
high-mass  supernovae to  escape, while  retaining the  ejecta  of SNe
whose progenitors were $\sim10$~$M_\sun$.  This could explain $\omega$
Cen's unique chemical evolution, as explained below.

Within  a similar  scenario, we  are forced  to look  for  stars which
produce  He  very  efficiently.   The  first  obvious  candidates  are
intermediate-mass  stars,  which according  to  various authors  (see,
e.g., Izzard  et al.\  2004) may indeed  produce large amounts  of He,
which   can   pollute   the   surrounding  nebula   during   the   AGB
phase. However, the amount of ejected He does not seem to be enough to
raise $Y$  to the  values needed to  explain the bMS.  Moreover, these
same stars should also produce C efficiently; the fact that we found a
similar C abundance for both the  rMS and bMS stars (see \S3) seems to
exclude this possibility.

Norris (2004) suggested that massive ($\sim 20$~$M_\sun$) stars can be
the source  of the high He  content. However, while  it is conceivable
that  heavier elements  collapse into  the central  black hole,  we do
expect that  a large amount  of ejected He  would be accompanied  by a
corresponding large amount of  CNO and other $\alpha$-elements such as
Mg and Si..   This fact seems to be  contradicted by the observational
evidence  described  in  \S  3.2.   An appealing  alternative  may  be
represented   by   the   smallest   among   core-collapse   supernovae
(SNe).  According  to  the  prescriptions by  Thielemann,  Nomoto,  \&
Hashimoto (1996), complemented  by data by Argast et  al.\ (2002), SNe
with  initial  masses  smaller  than  about  10--14~$M_\sun$\  produce
significant  amounts of  He,  while producing  only  small amounts  of
heavier  elements. According  to the  same  authors, the  ejecta of  a
10~$M_\sun$\ star might indeed have $\Delta Y/\Delta Z\sim 70$, enough
to explain the difference between the bMS and the rMS of $\omega$~Cen.
Among  the other  elements considered  in this  paper, we  notice that
models by Thielemann et al.\ (1996) predict that N is more abundant in
the ejecta of the less massive  core collapse SNe with respect to more
massive ones,  while C abundances increase  with increasing progenitor
mass; these predictions agree qualitatively with our observations.  On
the other  hand, there is no  prediction about Ba in  these SN models.
Ba is observed to be overabundant in metal-rich stars of $\omega$~Cen,
with  a pattern  characteristic of  the s-process  (see Smith  et al.\
2000).   AGB stars  are  supposed  to produce  most  of the  s-process
elements,  though it  is  possible  that massive  stars  make a  small
contribution.  Our data suggest a  moderate overabundance of Ba in the
bMS  stars;   this  fact  must  still  be   confronted  with  adequate
nucleosynthetic predictions.

There are two additional problems  which need to be taken into account
by any  model of the stellar  population history in  \omcen. The first
one, already raised by Norris (2004),  is that in order to elevate $Y$
from 0.24  to $\sim$0.38 one has to  assume that most, if  not all the
material from which  the bMS stars formed is made up  of the ejecta of
the  first generation  of stars.   The  second problem  is that  these
ejecta, which must  necessarily come from a large  number of stars (in
view of the  size of the bMS stellar population),  must have been well
homogenized in their metal content before the stars of the bMS formed,
as shown by the homogeneity found for the bMS in \S 4.

\acknowledgments

We thank  R.\ Kraft  and J.\ Norris  for useful discussions.   We also
thank the  anonymous referee whose  suggestions helped to  improve the
paper.   L.R.B.,  S.C.,  R.G.,   S.L.,  Y.M.,  and  G.P.\  acknowledge
financial support  by MIUR (PRIN2002).  J.A.\  and I.R.K.\ acknowledge
support by STScI grant GO 9444.


\begin{center}
\begin{table}[h!]
\caption{Observed bMS and rMS stars}
\begin{tabular}{|c|c|c|c|c|}
\hline
ID&RA(2000)&DEC(2000)&F606W&F606W-F814W\\
\hline
\ &\ &bMS&\ &\ \\
\hline
0530 &  13:25:26.605 &    -47:38:48.06 & 20.37 & 0.79\\
0577 &  13:25:26.945 &    -47:39:18.22 & 20.71 & 0.83\\
0718 &  13:25:28.177 &    -47:41:31.60 & 20.60 & 0.82\\
0795 &  13:25:27.834 &    -47:38:29.77 & 20.37 & 0.79\\
1416 &  13:25:31.581 &    -47:41:12.63 & 20.50 & 0.81\\
1616 &  13:25:32.108 &    -47:39:35.49 & 20.74 & 0.85\\
1993 &  13:25:33.688 &    -47:38:45.38 & 20.62 & 0.83\\
2522 &  13:25:36.322 &    -47:38:41.99 & 20.63 & 0.83\\
2529 &  13:25:37.045 &    -47:41:38.78 & 20.14 & 0.75\\
2739 &  13:25:37.772 &    -47:40:59.10 & 20.69 & 0.85\\
2822 &  13:25:37.952 &    -47:40:20.87 & 20.29 & 0.79\\
2938 &  13:25:38.026 &    -47:38:52.49 & 20.78 & 0.86\\
3256 &  13:25:39.718 &    -47:40:43.15 & 20.64 & 0.83\\
3348 &  13:25:40.197 &    -47:41:18.56 & 20.39 & 0.77\\
3977 & 13:25:42.130 &    -47:39:11.94 & 20.04 & 0.74\\
3990 &  13:25:41.996 &    -47:38:26.85 & 20.83 & 0.87\\
4085 &  13:25:42.438 &    -47:38:46.17 & 20.14 & 0.76\\
\hline
\ &\ &rMS&\ &\ \\
\hline
0179 &  13:25:25.223 &    -47:40:09.27 & 20.36 & 0.85\\
0237 &  13:25:25.767 &    -47:41:12.67 & 20.22 & 0.83\\
0568 &  13:25:27.303 &    -47:41:05.85 & 20.45 & 0.86\\
0571 &  13:25:27.259 &    -47:40:52.00 & 20.61 & 0.90\\
0664 &  13:25:27.526 &    -47:39:57.83 & 20.53 & 0.88\\
0851 &  13:25:28.477 &    -47:39:47.50 & 20.20 & 0.81\\
1259 &  13:25:30.480 &    -47:39:44.94 & 20.09 & 0.78\\
1350 &  13:25:30.625 &    -47:38:25.99 & 20.52 & 0.87\\
1475 &  13:25:31.389 &    -47:39:20.16 & 20.54 & 0.89\\
1677 &  13:25:32.794 &    -47:41:22.78 & 20.67 & 0.88\\
1980 &  13:25:33.925 &    -47:40:11.67 & 20.26 & 0.83\\
2294 &  13:25:35.745 &    -47:40:20.30 & 20.11 & 0.80\\
2900 &  13:25:38.425 &    -47:41:09.22 & 20.14 & 0.79\\
3080 &  13:25:38.871 &    -47:40:13.00 & 20.28 & 0.84\\
3222 &  13:25:39.779 &    -47:41:37.65 & 20.45 & 0.84\\
3533 &  13:25:40.434 &    -47:39:26.57 & 20.55 & 0.88\\
4302 &  13:25:43.376 &    -47:39:01.88 & 20.44 & 0.87\\
\hline
\end{tabular}
\end{table}
\end{center}


\begin{center}
\begin{table}
\caption{Spectral bands used for spectral indices}
\begin{tabular}{lcc}
\hline
Feature & Min. Wavelength & Max. Wavelength \\
          &    \AA          &      \AA        \\
\hline
Cont.      &   4051.0     &     4061.0      \\
H$_\delta$ &   4094.0     &     4112.0      \\ 
Ca~I~4227  &   4224.0     &     4230.0      \\ 
Cont.      &   4231.0     &     4236.0      \\
Cont.      &   4278.5     &     4286.0      \\ 
G-band     &   4294.5     &     4318.0      \\ 
H$_\gamma$ &   4334.5     &     4349.0      \\ 
Cont.      &   4356.0     &     4364.0      \\
Fe~I~4383  &   4380.0     &     4387.6      \\ 
Cont.      &   4440.0     &     4448.0      \\
\hline
\end{tabular}
\end{table}
\end{center}


\begin{center}
\begin{table}
\caption{Spreads in line indices}
\begin{tabular}{lccc}
\hline
           &   Expected &   bMS  &   rMS   \\
\hline
Fe I     &     0.031  &  0.033 &  0.063  \\
Ca I     &     0.050  &  0.061 &  0.071  \\
G-band   &     0.030  &  0.030 &  0.053  \\
\hline
\end{tabular}
\end{table}
\end{center}

\clearpage


\begin{figure}
\epsscale{1.00}
\plotone{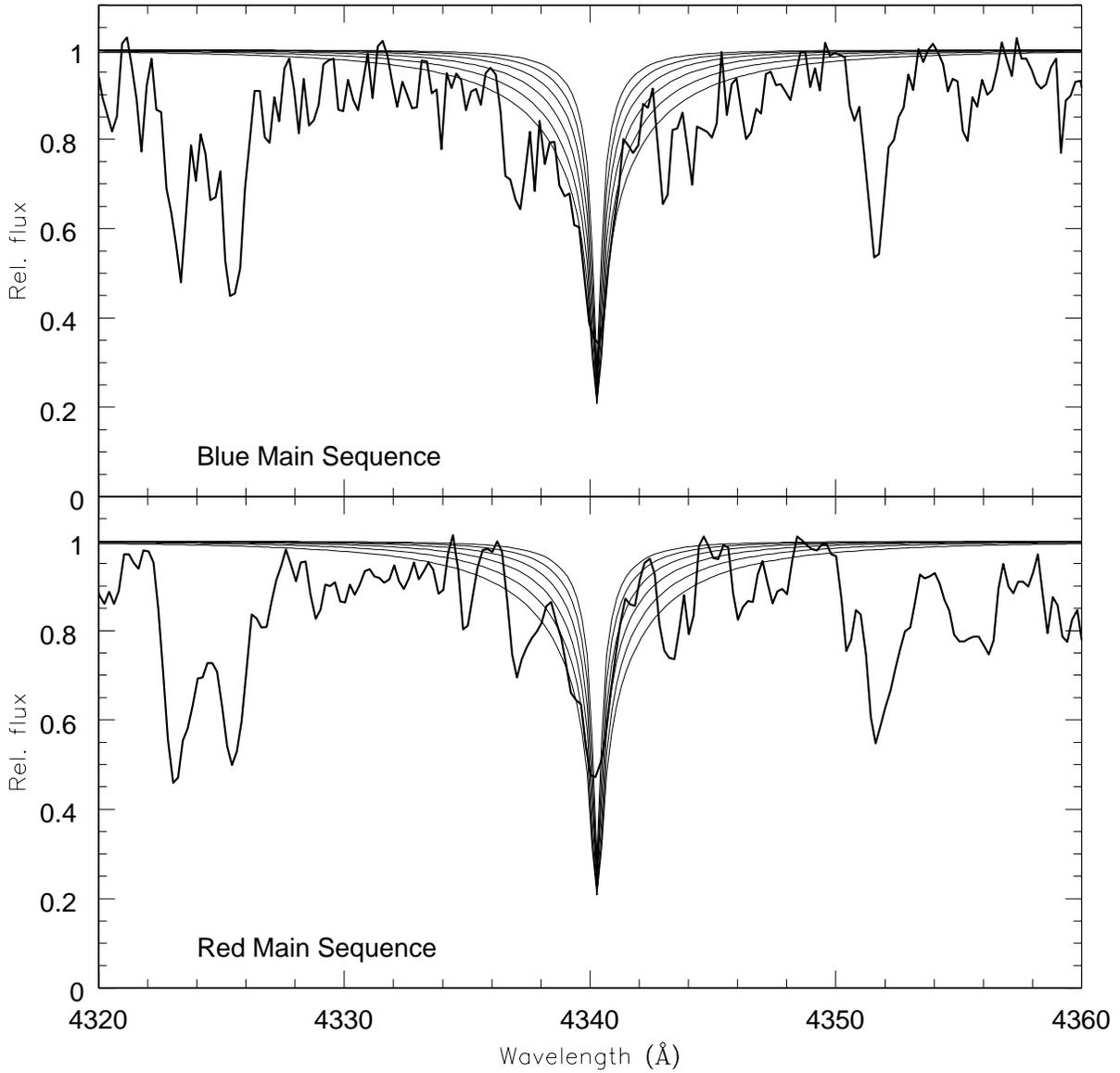}
\caption{The observed  H$_\gamma$ line in the average  bMS ({\it upper
panel})  and rMS  ({\it lower  panel})  spectra is  compared with  the
synthetic spectra for different  temperatures (5000, 5200, 5400, 5600,
5800, and 6000  K). Gravities and metal abundances  in the models were
chosen to be compatible with the final adopted values. }
\end{figure}

\clearpage


\begin{figure}
\epsscale{1.00}
\plotone{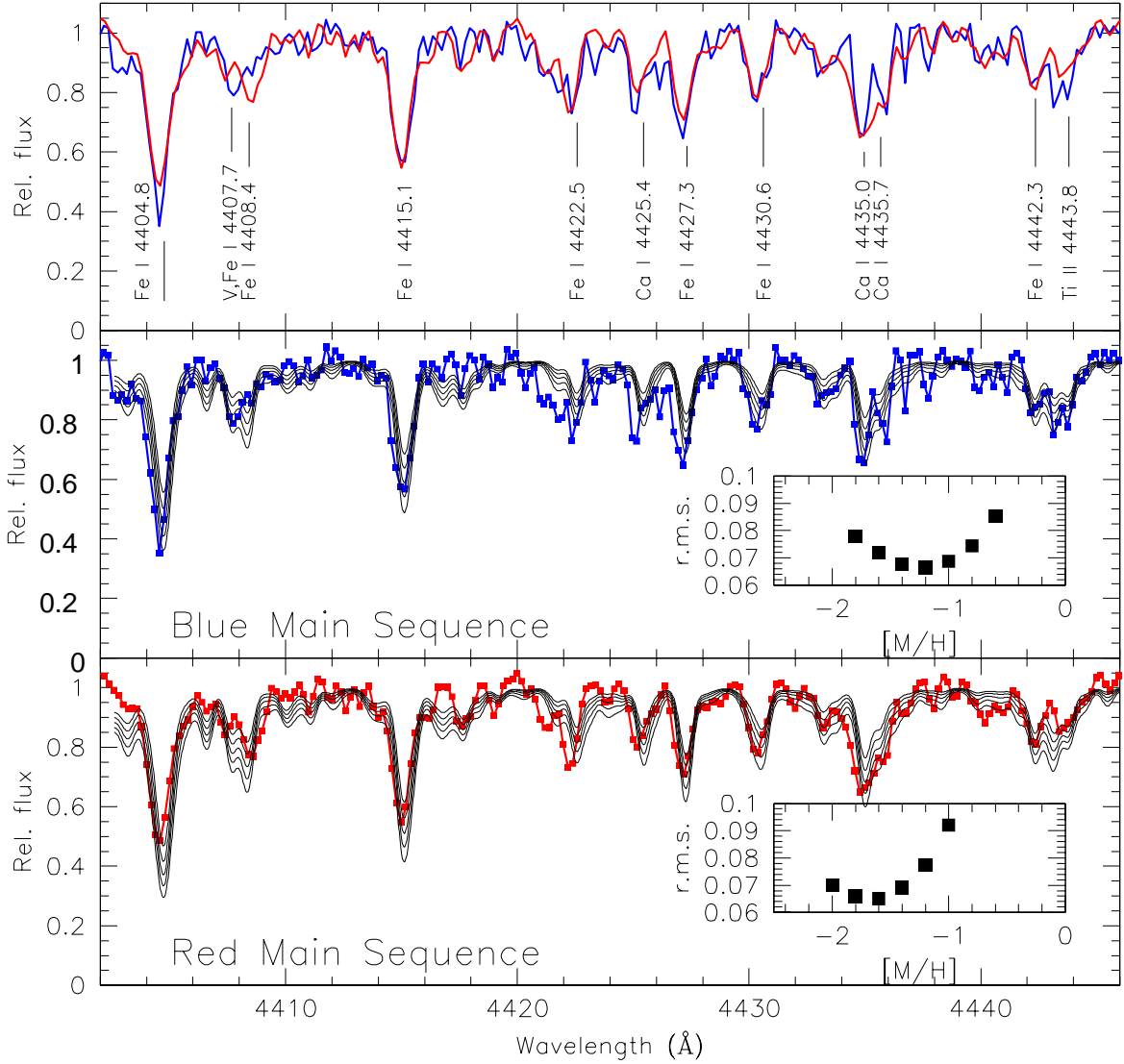}
\caption{The average  bMS ({\it blue  line}) and rMS ({\it  red line})
spectra are overplotted in the {\it upper panel}, where a few relevant
spectral   lines   are   also   indicated.    Though   a   few   lines
(e.g.  Fe~I~4404.8, Ti~II~4443.8)  are  clearly different  in the  two
spectra,  the effect on  the line  strength due  to the  difference in
metal content between  the bMS and rMS stars  is partially compensated
by  the temperature  difference (see  also Fig.~6).   The  average bMS
({\it middle panel})  and the average rMS ({\it  lower panel}) spectra
are compared with synthetic spectra for different metallicities ([M/H]
=  $-$1.6, $-$1.4, $-$1.2,  and $-$1.0  for the  bMS; [M/H]  = $-$2.0,
$-$1.8, $-$1.6,  $-$1.4, and  $-$1.2 for the  rMS). In these  two last
panels, observed  spectra are represented  by colored lines  and dots;
synthetic spectra are the thin solid lines.}
\end{figure}

\clearpage


\begin{figure}
\epsscale{1.00}
\plotone{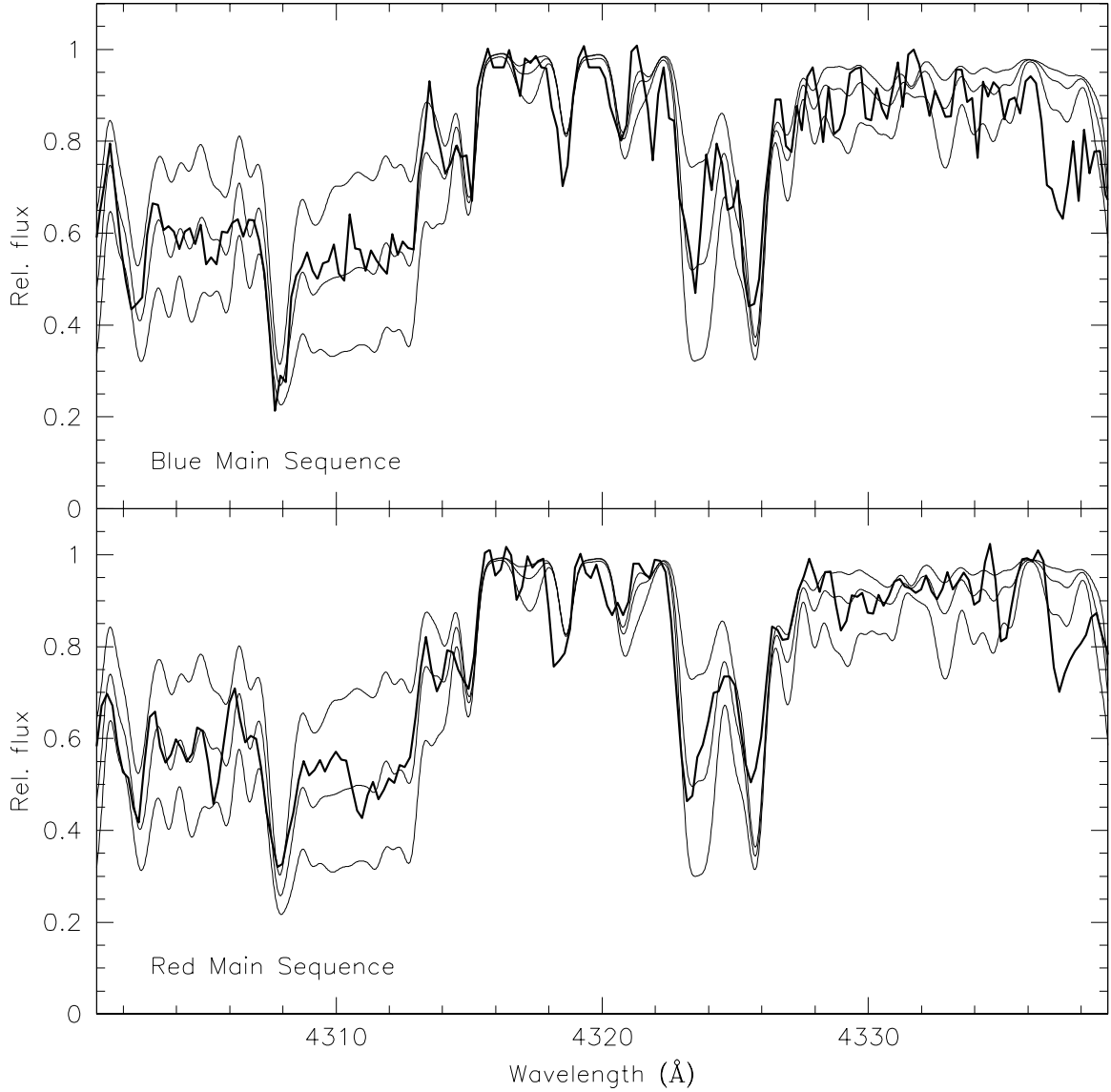}
\caption{The average bMS ({\it upper panel}) and the average rMS ({\it
lower  panel}) spectra  are  compared with  synthetic  spectra in  the
region    4301--4339~\AA,   including    the    band-heads   of    the
CH-band.  Synthetic spectra were  computed for  atmospheric parameters
appropriate for the stars, and C  abundances of [C/M] = $-0.5$, 0, and
+0.5 dex. Thick  solid lines are the observed  spectra, thin lines are
the synthetic spectra.}
\end{figure}

\clearpage


\begin{figure}
\epsscale{1.00}
\plotone{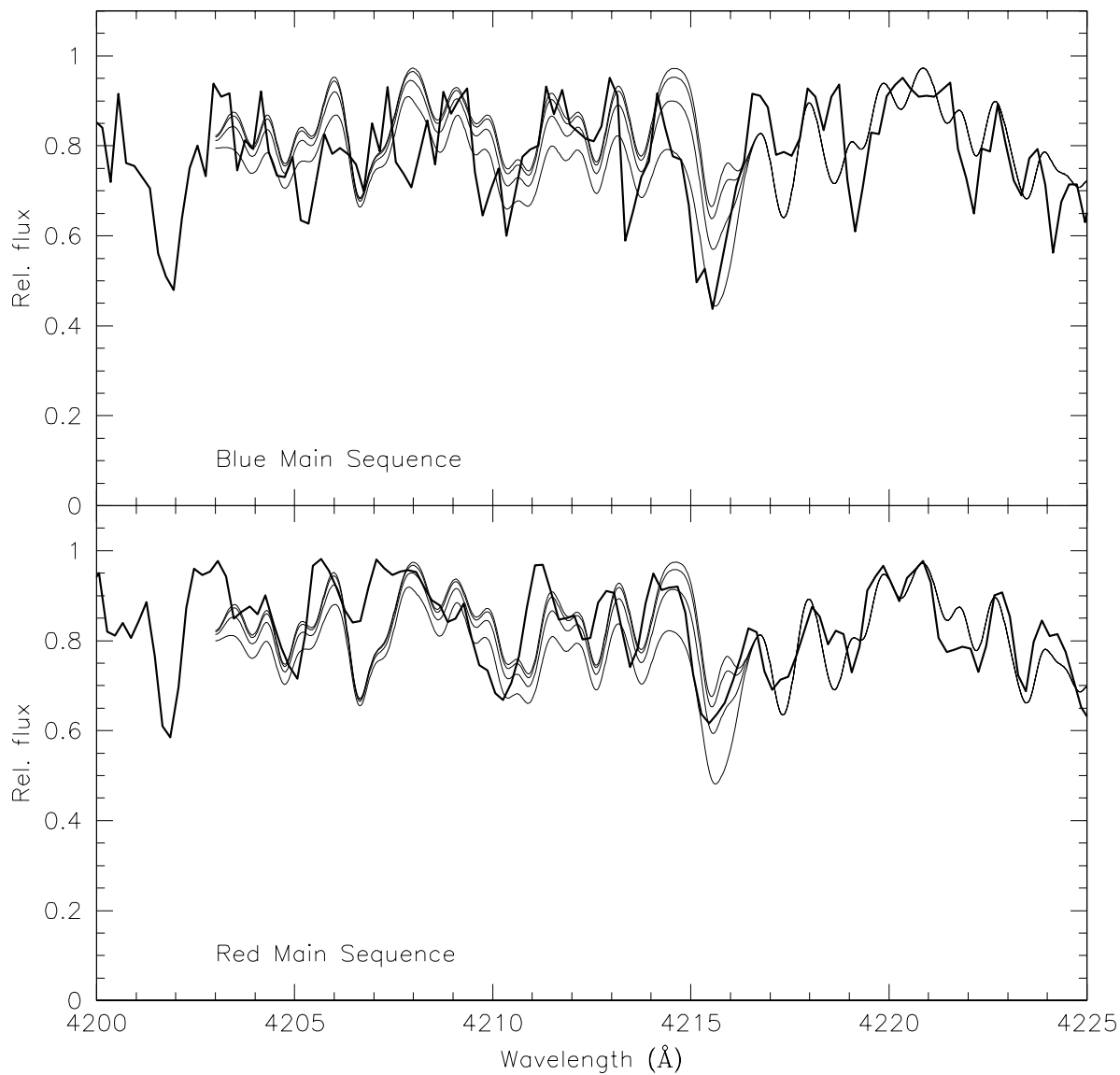}
\caption{The average bMS ({\it upper panel}) and the average rMS ({\it
lower  panel}) spectra  are  compared with  synthetic  spectra in  the
region 4200--4225~\AA,  including the band-heads  of the $\Delta  v$ =
2--0 violet  CN band. Synthetic spectra were  computed for atmospheric
parameters appropriate for  the stars, [C/M] = 0,  and N abundances of
[N/M] = 0,  0.5, 1.0, and 1.5 dex. Thick solid  lines are the observed
spectra, thin lines are the synthetic spectra.}
\end{figure}

\clearpage


\begin{figure}
\epsscale{1.00}
\plotone{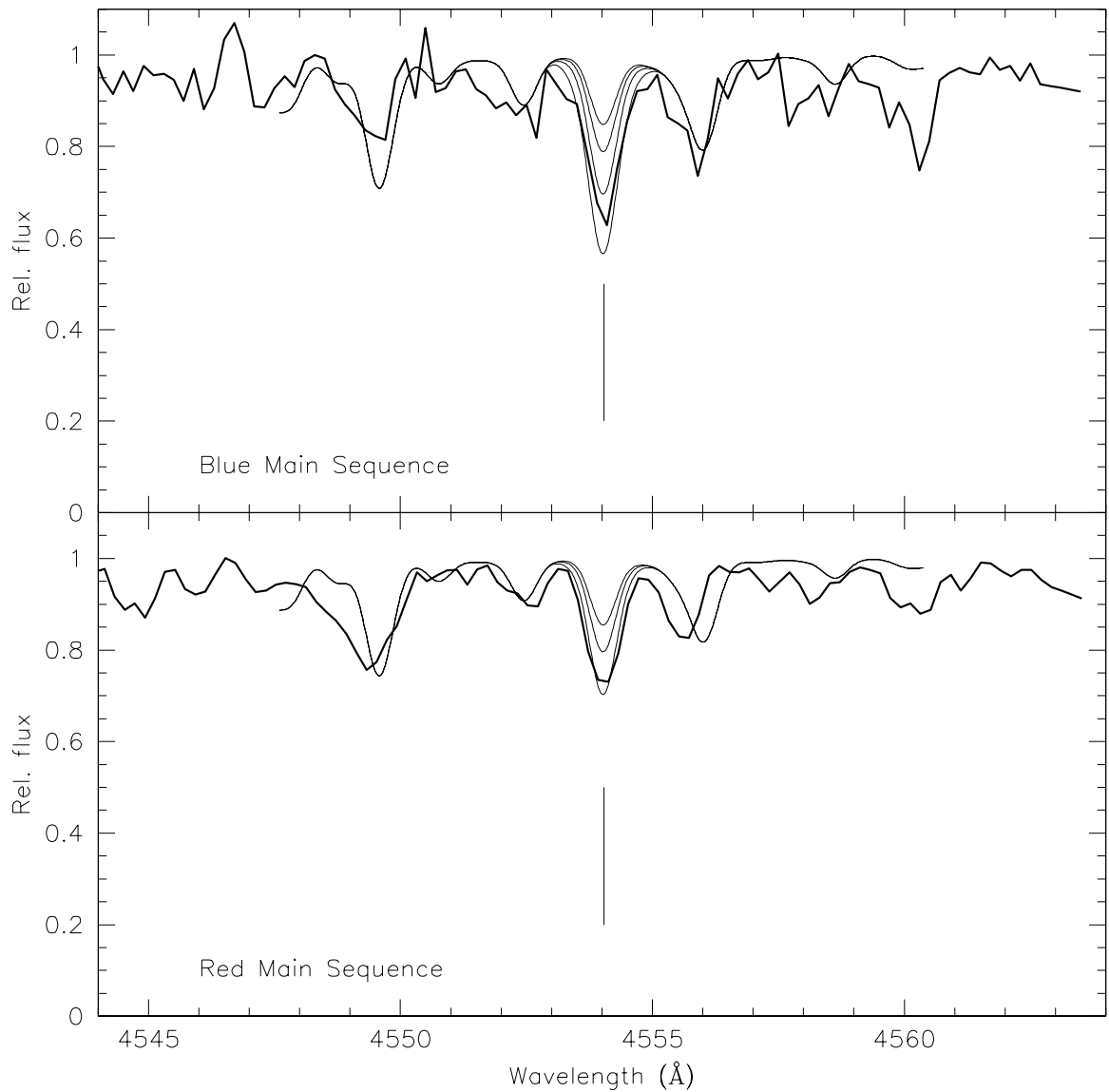}
\caption{The average bMS ({\it upper panel}) and the average rMS ({\it
lower  panel}) spectra  are  compared with  synthetic  spectra in  the
region  of  4544--4564~\AA, including  the  resonance  Ba  II line  at
4554~\AA. Synthetic  spectra were computed  for atmospheric parameters
appropriate for  the stars, and Ba  abundances of [Ba/M]  = $-0.5$, 0,
0.5, and  1.0 (upper panel),  and [Ba/M] =  $-0.5$, 0, and  0.5 (lower
panel). Thick solid lines are the observed spectra, thin lines are the
synthetic spectra.  }
\end{figure}

\clearpage


\begin{figure}
\epsscale{1.00}
\plotone{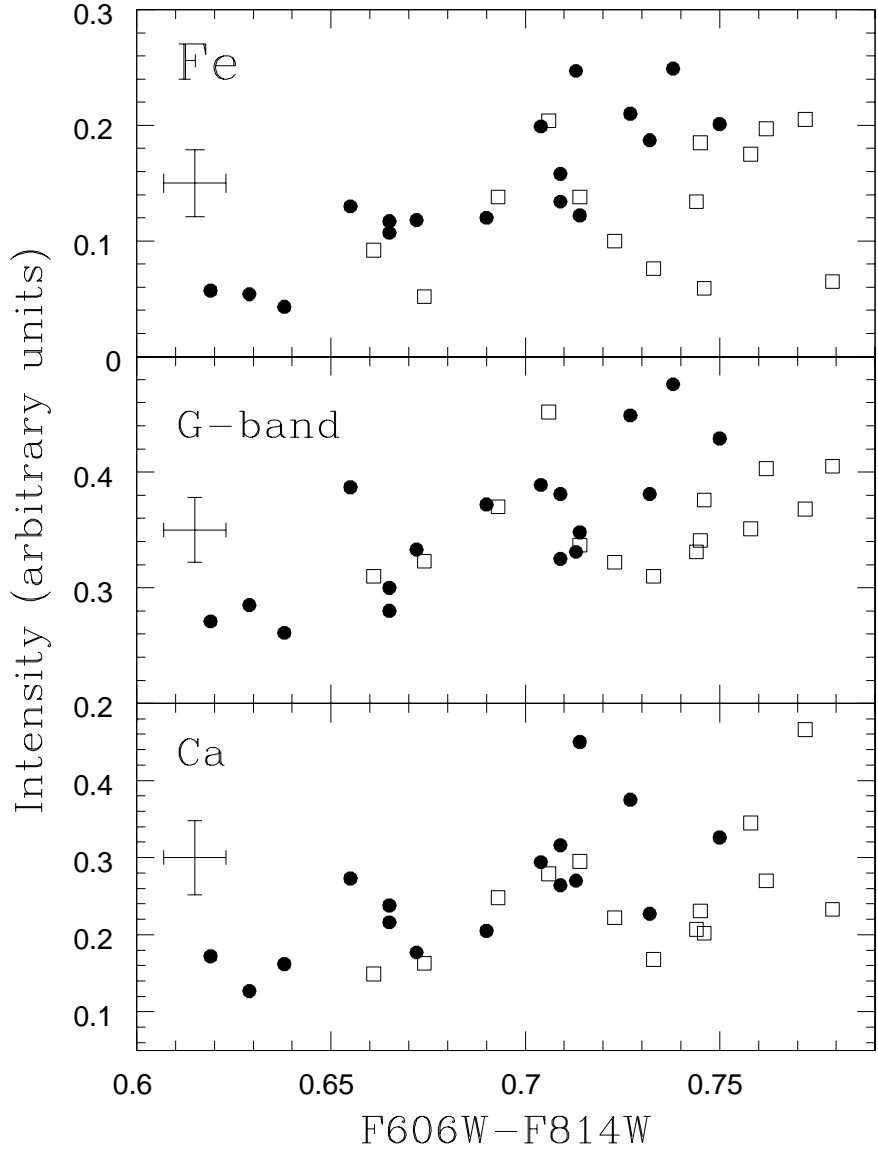}
\caption{Runs  of   various  spectral  indices   ({\it  upper  panel:}
Fe~I~4383;  {\it  middle  panel:}   G  band;  and  {\it  lower  panel}
Ca~I~4227)   with  the  F606W   $-$  F814W   color  for   the  program
stars.   Filled   and   open   symbols   are  bMS   and   rMS   stars,
respectively. Internal error bars are also shown for comparison.}
\end{figure}

\clearpage


\begin{figure}
\plotone{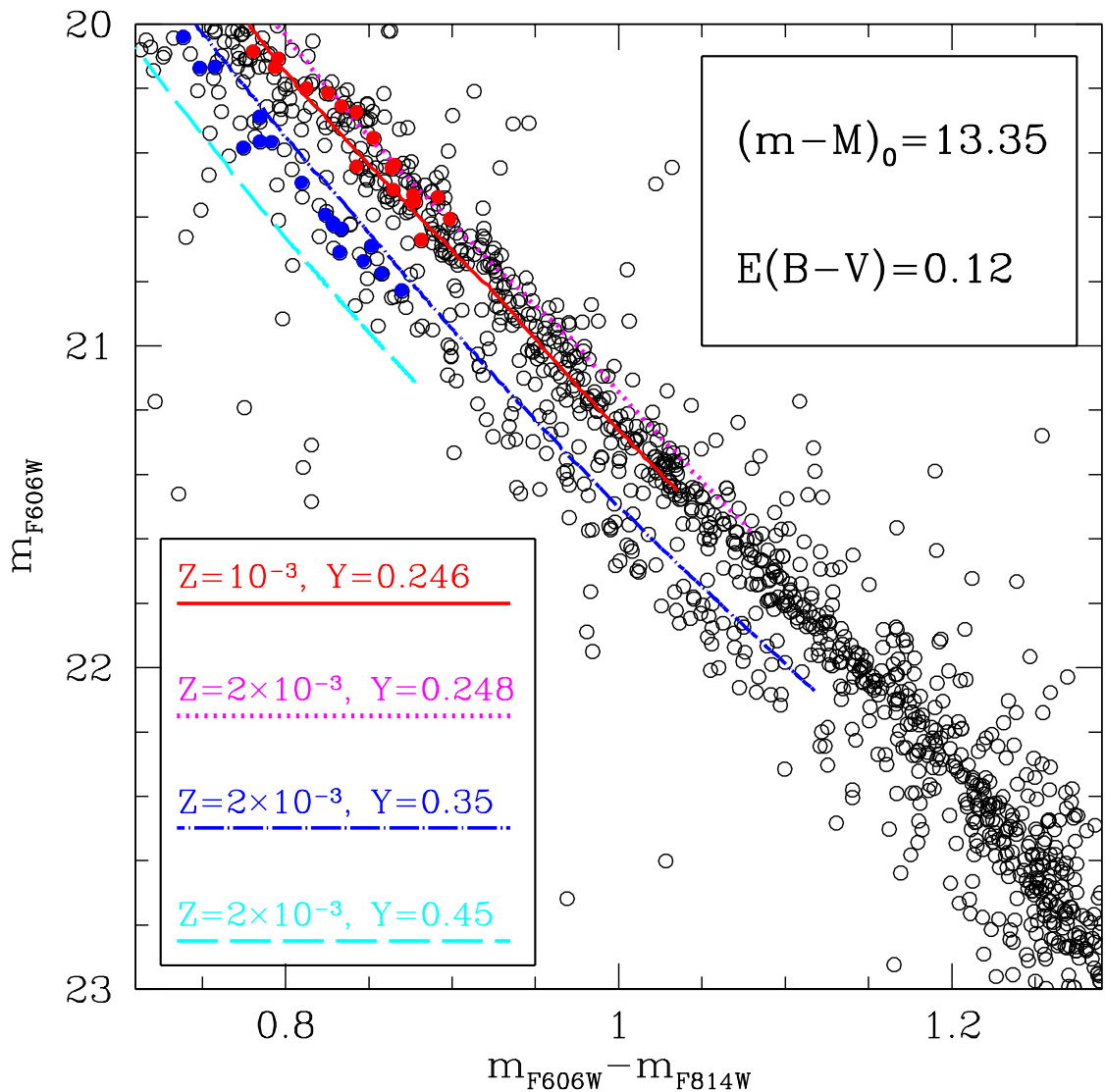}
\caption{Comparison of  the ACS  CMD of \omcen\  (calibrated following
Bedin et  al.\ 2004b) with  isochrones calculated for  the metallicity
determined  in  Fig.~1.  The  bMS  can  be  reproduced  only  assuming
$Y>0.35$.  The blue and red filled dots show the bMS and rMS stars for
which we collected our spectra.}
\end{figure}

\end{document}